\documentclass{elsart5p}

\journal{Phys. Lett. B}

\usepackage{latexsym}
\usepackage{epsfig}

\usepackage{amssymb}

\usepackage{amsmath}

\usepackage{color}

\begin{document}

\begin{frontmatter}

\title{Classicalization of Quantum Fluctuations at the Planck Scale \\ 
in the $R_{\rm h}=ct$ Universe}

\author{Fulvio Melia\ead{fmelia@email.arizona.edu}}

\address{Department of Physics, the Applied Math Program, and Department of Astronomy, \\
The University of Arizona, Tucson AZ 85721}

\begin{abstract}
The quantum to classical transition of fluctuations in the early universe
is still not completely understood. Some headway has been made incorporating the effects
of decoherence and the squeezing of states, though the methods and procedures continue
to be challenged. But new developments in the analysis of the most recent {\it Planck} data
suggest that the primordial power spectrum has a cutoff associated with the
very first quantum fluctuation to have emerged into the semi-classical universe from the
Planck domain at about the Planck time. In this paper, we examine the implications of
this result on the question of classicalization, and demonstrate that the birth of
quantum fluctuations at the Planck scale would have been a `process' supplanting the
need for a `measurement' in quantum mechanics. Emerging with a single wavenumber, these
fluctuations would have avoided the interference between different degrees of freedom
in a superposed state. Moreover, the implied scalar-field potential had an equation-of-state
consistent with the zero active mass condition in general relativity, allowing the
quantum fluctuations to emerge in their ground state with a time-independent frequency.
They were therefore effectively quantum harmonic oscillators with classical correlations
in phase space from the very beginning.
\end{abstract}

\begin{keyword} cosmology \sep quantum fluctuations \sep inflation 
\PACS 04.20.Ex \sep 95.36.+x \sep 98.80.-k \sep 98.80.Jk
\end{keyword}
\end{frontmatter}

\section{Introduction}
In standard inflationary $\Lambda$CDM cosmology, the early universe underwent a phase of
quasi-exponential expansion due to the action of a scalar (inflaton) field with a near-flat
potential, a process typically referred to as `slow-roll' inflation
\cite{Starobinsky:1980,Guth:1981,Linde:1982,Albrecht:1982,Linde:1983}.

The concept of inflation was introduced to solve several global inconsistencies with 
standard Big Bang cosmology, notably its horizon problem, but an even more important 
consequence of inflation was realized with the introduction of quantum effects 
\cite{Mukhanov:1981,Mukhanov:1982,Starobinsky:1982,Guth:1982,Hawking:1982}. Small 
inhomogeneous fluctuations on top of an otherwise isotropic and homogeneous background 
are now broadly believed to be the explanation for the anisotropies in the cosmic 
microwave background (CMB) and the formation of large-scale structure 
\cite{Bardeen:1983}. Beginning as quantum seeds in the distant conformal past, well before 
the Planck time $t_{\rm Pl}$, these inhomogeneities grew with the expansion of the universe 
and produced a near scale-free power spectrum as they crossed the Hubble horizon 
\cite{Stewart:1993}.  Indeed, the small departure from a perfectly scale-free distribution 
is viewed (in this model) as a consequence of `Hubble friction,' whose strength and duration 
are directly attributable to the detailed shape of the inflaton potential. Considerable work 
has been carried out over the past four decades to use this framework in order to constrain 
the physical conditions in the early universe 
\cite{Martin:2006,Lorenz:2008,Larson:2011,Komatsu:2011,Melia:2020b}, at energies up to the 
grand unified scale at $\sim 10^{15}$ GeV. 

But the CMB anisotropies and galaxy clusters are classical, so we are faced with the
problem of understanding how quantum fluctuations in the inflaton field transitioned
into purely classical objects---an issue closely related to the long-standing
`measurement' problem in quantum mechanics, i.e., how does a deterministic outcome appear
in a measurement process performed on a quantum system prepared in a superposed state?
Different authors use a range of descriptions to characterize the distinctions between
quantum and classical states (see, e.g., refs.~\cite{Nambu:1992,Mijic:1997,Lee:2001})
but, at a fundamental level, a successful transition from the former to the latter requires 
the substantial elimination of interference between the various degrees of freedom in a 
superposed quantum state and the subsequent appearance of macroscopic variables. The 
second condition is the emergence of classical correlations in the phase space of such 
canonical variables. In other words, classical trajectories (e.g., orbits) with well-defined 
values of these variables need to be established.

The age-old (mostly heuristic) Copenhagen interpretation suggests that a quantum system
remains in its superposed state until an observer performs a measurement on it, the
interaction of which causes it to `collapse' into one of the eigenstates of the
observable being measured \cite{Zurek:1981}. The original version of this 
concept has since been refined with the phenomenon of decoherence, in which the system 
becomes entangled with its environment \cite{Joos:1985}, consisting of a 
very large number of degrees of freedom, bringing the original quantum state into an 
ensemble of classical looking ones. The manifestation of this problem specifically in
the cosmological context has sometimes been framed in the sense that the CMB anisotropies 
constitute a measurement of the field variable \cite{Kiefer:1998,Kiefer:2007,Kiefer:2009}.

The formalism based on decoherence, however, is subject to considerable debate.
It has been argued by some \cite{Adler:2003,Schlosshauer:2004,Sudarsky:2011}
that decoherence by itself does not solve the problem of a single outcome. As we shall
discuss later in this paper, this mechanism is arguably even less likely to be
relevant to the early universe---a `closed' system where the distinction between
the quantum state and an `environment' is essentially nonexistent. Other types
of approaches, such as Bohmian mechanics \cite{Bohm:1952a,Bohm:1952b}, do not by
themselves predict testable features that may be verified or refuted.

Our entry point into this discussion is motivated by several significant new developments
in the updated analysis of the latest {\it Planck} data release \cite{Planck:2018},
which build on many attempts made over the past two decades to find specific features
associated with the primordial power spectrum, ${\mathcal{P}}(k)$. For example, several large-angle 
anomalies have been present in the CMB maps since the 1990's. The most recent studies
\cite{MeliaLopez:2018,Melia:2020a} of the {\it Planck} data have extended our ability 
to examine whether these issues are more likely due to instrumental or other systematic 
effects, or whether they truly represent real characteristics in ${\mathcal{P}}(k)$. And
in \S~3.1, we shall briefly discuss why these anomalies may now be viewed as possibly being
due to a cutoff $k_{\rm min}$ in the power spectrum. Its value, however, calls into question 
whether slow-roll inflation remains viable \cite{LiuMelia:2020}, at least based on the 
inflaton potentials proposed thus far. As we shall explore in \S~3.2, a more interesting 
interpretation for the appearance of $k_{\rm min}$ is that this represents the first mode 
to exit from the Planck scale into the semi-classical universe at about the Planck time. 
If correct, this interpretation would clearly have a considerable impact on the 
quantum-to-classical transition.

In this paper, we focus on how this interpretation could alter our view of the long-standing
classicalization problem in the early universe. The conventional view has been that quantum
fluctuations were seeded in the Bunch-Davies vacuum, though without much guidance other than
they emerged in the ground state. But if they appeared at the Planck scale at about the Planck
time, the manner in which the quantum states were `prepared' may have some bearing on the
classicalization problem. In \S~2, we shall summarize the current status with the quantum-to-classical
transition of fluctuations in the inflaton field, and then transition into a discussion of the
analogous situation for a non-inflationary scalar field with the zero active mass equation-of-state
(relevant to the above interpretation of $k_{\rm min}$) in \S~3. We shall describe
the impact of these new developments on the classicalization question in \S~4 and end with
our conclusions in \S~V.

\section{Classicalization of Quantum Fluctuations in the Inflaton Field}
\subsection{Quantum Fluctuations in the Inflaton Field}
The essential steps for deriving the perturbation growth equation are by now well known,
and one may find many accounts of this procedure in both the primary and secondary
literature. For the sake of brevity, we here show only some key results---chiefly those
that will also be relevant to our discussion of quantum fluctuations at the Planck scale
in \S\S~3 and 4---and refer the reader to several other influential publications for
all the details.

The perturbed Friedmann-Lema{\^i}tre-Robertson-Walker (FLRW) spacetime for linearized
scalar fluctuations is given by the line element
\cite{Bardeen:1980,Kodama:1984,Mukhanov:1992,Bassett:2006}
\begin{eqnarray}
ds^2 &=& (1+2A)\,dt^2-2a(t)(\partial_iB)\,dt\,dx^i-\nonumber\\
&\null&\hskip-0.1in a^2(t)\left[(1-2\psi)\delta_{ij}+2(\partial_i
\partial_jE)+h_{ij}\right]\,dx^i\,dx^j\,,\quad
\label{eqn:metric}
\end{eqnarray}
where indices $i$ and $j$ denote spatial coordinates, $a(t)$ is the expansion
factor, and $A$, $B$, $\psi$ and $E$ describe the scalar metric perturbations,
while $h_{ij}$ are the tensor perturbations.

For small perturbations about the homogeneous scalar field $\phi_0(t)$,
\begin{equation}
\phi(t,\vec{x}) = \phi_0(t)+\delta\phi(t,\vec{x})\;,
\end{equation}
one can identify the curvature perturbation ${\mathcal{R}}$ on hypersurfaces orthogonal to worldlines
in the comoving frame written as a gauge invariant combination of the scalar field perturbation
$\delta\phi$ and the metric perturbation $\psi$ \cite{Bardeen:1980}:
\begin{equation}
{\mathcal{R}}\equiv \psi+\left({H\over \dot{\phi}}\right)\,\delta\phi\;.
\end{equation}
A solution for ${\mathcal{R}}$ may then be obtained by (i) using an expansion in Fourier modes,
\begin{equation}
{\mathcal{R}}(\tau,{\bf x})\equiv\int {d^3{\bf k}\over (2\pi)^{3/2}}\;{\mathcal{R}}_{\bf k}(\tau)
e^{i{\bf k}\cdot{\bf x}}\;,\label{eqn:Fourier}
\end{equation}
and (ii) inserting the linearized metric (Eq.~\ref{eqn:metric}) into Einstein's equations,
which together yield the following perturbed equation of motion for each mode {\bf k}:
\begin{equation}
{\mathcal{R}}_{\bf k}^{\prime\prime}+2\left({z^\prime\over z}\right){\mathcal{R}}_{\bf k}^\prime+
k^2{\mathcal{R}}_{\bf k}=0\;.
\label{eqn:mathcalR}
\end{equation}
Overprime denotes a derivative with respect to conformal time $d\tau=dt/a(t)$, and
we have defined the variable
\begin{equation}
z\equiv {a(t)(\rho_\phi+p_\phi)^{1/2}\over H}\;.
\label{eqn:z}
\end{equation}

Using the canonically normalized Mukhanov-Sasaki variable \cite{Kodama:1984,Mukhanov:1992}
\begin{equation}
u_{\bf k}\equiv z{\mathcal{R}}_{\bf k}\;,
\label{eqn:MS}
\end{equation}
Equation~(\ref{eqn:mathcalR}) may be recast into the more familiar form of a parametric oscillator
with a time-dependent frequency, known (in this context) as the Mukhanov-Sasaki equation,
\begin{equation}
u_{\bf k}^{\prime\prime}+\omega_{\bf k}^2(\tau)\,u_{\bf k}=0\;,
\label{eqn:uk}
\end{equation}
where the time-dependent frequency is
\begin{equation}
\omega_{\bf k}(\tau)^2\equiv k^2-{z^{\prime\prime}\over z}\;.
\label{eqn:freq}
\end{equation}
In Minkowski spacetime, the scale factor $a(\tau)$ is constant and $\omega_{\bf k}=k$, which
reduces Equation~(\ref{eqn:uk}) to that of the more basic harmonic oscillator with a
constant frequency.

Since $\omega_{\bf k}(\tau)$ depends solely on $\tau$ and $k$ (not the direction of {\bf k}),
the most general solution to the Mukhanov-Sasaki equation may be written
\begin{equation}
u_{\bf k}=a_{\bf k}^-u_k(\tau)+a_{\bf -k}^+u_k^*(\tau)\;,
\end{equation}
where $u_k(\tau)$ and its complex conjugate are linearly independent solutions to Equation~(\ref{eqn:uk}),
and are the same for all Fourier modes with $k=|{\bf k}|$. In general, however, the integration
constants $a_{\bf k}^-$ and $a_{\bf -k}^+$ may depend on the direction of {\bf k}. From
Equations~(\ref{eqn:Fourier}) and (\ref{eqn:MS}), one may thus write the complete solution to
the Mukhanov-Sasaki equation as
\begin{equation}
u(\tau,{\bf x})=\int {d^3{\bf k}\over (2\pi)^{3/2}}\left[a_{\bf k}^-
u_k(\tau)e^{i{\bf k}\cdot{\bf x}}+a_{\bf k}^+u_k^*(\tau)e^{-i{\bf k}\cdot{\bf x}}\right],\quad
\label{eqn:utotal}
\end{equation}
which is manifestly real since $a_{\bf k}^+=(a_{\bf k}^-)^*$.

Our derivation of Equation~(\ref{eqn:utotal}) has made no reference to the definite size
of the inhomogeneities. This result therefore merely shows us the dynamics of such
inhomogeneities if they came into existence. To address their presence and magnitude, one
must consider states at the quantum level, where the fluctuation field becomes an operator.
The next step is therefore the quantization of the field, noting that the Hilbert space for
the quantum field will be a direct product of individual Hilbert spaces for the Fourier modes.

The field $u(\tau,{\bf x})$ is quantized like the harmonic oscillator, except that the frequency
(Eq.~\ref{eqn:freq}) is now time dependent, which is itself a consequence of the curved spacetime
within which these modes are evolving. To properly normalize the quantum fluctuation, one must
therefore choose the vacuum carefully, since the spacetime is generally not Minkowskian.

The field $u(\tau,{\bf x})$ and its canonical conjugate `momentum' $\pi(\tau,{\bf x})\equiv 
u^\prime(\tau,{\bf x})$ are promoted to quantum operators $\hat{u}$ and $\hat{\pi}$, satisfying
the standard equal-time canonical quantization relations
\begin{eqnarray}
&\null&[\hat{u}(\tau,{\bf x}_1),\hat{\pi}(\tau,{\bf x}_2)]=i\delta^3({\bf x}_1-{\bf x}_2)\nonumber\\
&\null&[\hat{u}(\tau,{\bf x}_1),\hat{u}(\tau,{\bf x}_2)]=0\nonumber\\
&\null&[\hat{\pi}(\tau,{\bf x}_1),\hat{\pi}(\tau,{\bf x}_2)]=0\;.
\label{eqn:can}
\end{eqnarray}
The constants of integration $a_{\bf k}^-$ and $a_{\bf k}^+$ in the mode expansion of
$u(\tau,{\bf x})$ become operators $\hat{a}_{\bf k}^-$ and $\hat{a}_{\bf k}^+$, so that
\begin{equation}
\hat{u}(\tau,{\bf x})=\int {d^3{\bf k}\over (2\pi)^{3/2}}\left[\hat{a}_{\bf k}^-
u_k(\tau)e^{i{\bf k}\cdot{\bf x}}+\hat{a}_{\bf k}^+u_k^*(\tau)e^{-i{\bf k}\cdot{\bf x}}\right].\quad
\label{eqn:fieldop}
\end{equation}
We thus see from Equations~(\ref{eqn:can}) and (\ref{eqn:fieldop}) that
\begin{eqnarray}
&\null&[\hat{a}^-_{\bf k},\hat{a}^+_{{\bf k}^\prime}]=\delta^3({\bf k}-{\bf k}^\prime)\nonumber\\
&\null&[\hat{a}^-_{\bf k},\hat{a}^-_{{\bf k}^\prime}]=0\nonumber\\
&\null&[\hat{a}^+_{\bf k},\hat{a}^+_{{\bf k}^\prime}]=0\,,
\label{eqn:raiselow}
\end{eqnarray}
which allows us to interpret $\hat{a}_{\bf k}^-$ and $\hat{a}_{\bf k}^+$ as annihilation and creation
operators, respectively, creating and annihilating excitations, or particles, of the field
$u(\tau,{\bf x})$. And in the usual fashion, the quantum states in the Hilbert space are constructed
by a repeated application of the creation operator,
\begin{equation}
|n_{{\bf k}_1},n_{{\bf k}_2},\cdots\rangle = {1\over\sqrt{n_{{\bf k}_1}!\,n_{{\bf k}_2}!\cdots}}
\left\{(\hat{a}_{{\bf k}_1}^+)^{n_{{\bf k}_1}}(\hat{a}_{{\bf k}_2}^+)^{n_{{\bf k}_2}}\cdots\right\}|0\rangle,\;
\label{eqn:Hilbertstate}
\end{equation}
starting with the vacuum state $|0\rangle$, which is defined by
\begin{equation}
\hat{a}_{\bf k}^-|0\rangle=0\;.
\end{equation}

In de Sitter space, where $a(\tau)=-(H\tau)^{-1}$ (in terms of the Hubble constant $H$ during
inflation), the exact solution to Equation~(\ref{eqn:uk}) is
\begin{equation}
u_k(\tau)=A_k{e^{-ik\tau}\over\sqrt{2k}}\left(1-{i\over k\tau}\right)+
          B_k{e^{ik\tau}\over\sqrt{2k}}\left(1+{i\over k\tau}\right)\;,
\label{eqn:uksoln}
\end{equation}
where $A_k$ and $B_k$ are integration constants to be fixed by the choice of vacuum. In a
time-independent spacetime, the normalization of the wavefunction is obtained by imposing
canonical quantization or, equivalently, by minimizing the expectation value of the
Hamiltonian in the vacuum state. For example, in Minkowski space, one would find from
Equation~(\ref{eqn:uk}) with $\omega_k=k$, and the Hamiltonian written in terms of
$u_k$ and $\pi_k$, that
\begin{equation}
u_k(\tau)={e^{-ik\tau}\over\sqrt{2k}}\;\;{\rm (Minkowski)}\;.
\label{eqn:Mink}
\end{equation}
In de Sitter spacetime, such a solution would therefore correspond to $|\tau|\rightarrow\infty$,
and $A_k=1$, $B_k=0$. This is interpreted to mean that in the remote conformal past, all
modes of current interest were much smaller than the Hubble radius, allowing us to ignore
curvature effects on the mode normalization, thereby defining a unique physical vacuum known
as the Bunch-Davies vacuum \cite{Bunch:1978}.

With the time evolution
\begin{equation}
u_k(\tau)={e^{-ik\tau}\over\sqrt{2k}}\left(1-{i\over k\tau}\right)\;\;{\rm (de\; Sitter)}\;,
\end{equation}
Equation~(\ref{eqn:fieldop}) gives us the full form of the field operator $\hat{u}(\tau,{\bf x})$,
which may be used to find the quantum states of the inflaton scalar field. As we shall
see in \S~3, much of this derivation remains intact for the (non-inflationary) numen field as
well, though with several crucial differences we shall discuss shortly. (As we shall 
see in \S~3.2, it will be necessary to clearly distinguish between inflaton scalar fields and
those that do not produce an inflated expansion. We shall therefore informally refer to the latter
as `numen' fields.)

We now turn our attention to describing the vacuum state with a wavefunctional that gives the probability
amplitudes for different field configurations in Fourier space. Though we have developed the expression 
for the field operator, $\hat{u}(\tau,{\bf x})$, in the Heisenberg picture, finding the various occupation
numbers in a quantum state such as Equation~(\ref{eqn:Hilbertstate}) is more easily accomplished
using a functional approach in the Schr\"odinger picture (see, e.g., refs.~\cite{Polarski:1996,Martin:2008}).

As long as the various $k$-modes are all independent, the Hilbert space for the field operator
$\hat{u}(\tau,{\bf x})$, containing eigenstates $|u\rangle$, is simply a direct product of Hilbert
spaces for the different Fourier components. If a state is initially decomposed as $|\Psi\rangle$,
the probability amplitude of measuring the field configuration $u(\tau,{\bf x})$ is then the
wave functional $\Psi[u(\tau,{\bf x})]\equiv \langle u|\Psi\rangle$ where, in the Schr\"odinger
approach, one may factor it into independent mode components according to:
\begin{equation}
\Psi[u(\tau,{\bf x})]=\prod_k\Psi_{\bf k}[u_k(\tau)]\;,
\label{eqn:Psitotal}
\end{equation}
each of which is a solution to Schr\"odinger's equation:
\begin{equation}
i{\partial\Psi_{\bf k}\over\partial\tau}=\hat{H}_{\bf k}\Psi_{\bf k}\;.
\label{eqn:Schro}
\end{equation}
The Hamilton operator in Fourier space may be written
\begin{equation}
\hat{H}={1\over 2}\int d^3{\bf k}\,\hat{H}_{\bf k}=
{1\over 2}\int d^3{\bf k}\left[\hat{\pi}_{\bf k}\hat{\pi}_{\bf k}^\dag+
\omega_{\bf k}^2\hat{u}_{\bf k}\hat{u}_{\bf k}^\dag\right]\;.
\label{eqn:Ham}
\end{equation}

The solution to Equations~(\ref{eqn:Schro}) and (\ref{eqn:Ham}) for a harmonic oscillator is
well known, and has the form of a Gaussian function:
\begin{equation}
\Psi_{\bf k}[u_{\rm k}(\tau)]=N_{\bf k}(\tau)e^{-\Omega_{\bf k}(\tau)(u_{\bf k})^2}\;,
\label{eqn:Psi}
\end{equation}
where
\begin{equation}
|N_{\bf k}|\equiv \left({2{\rm Re}\{\Omega_{\bf k}\}\over\pi}\right)^{1/4}
\end{equation}
is the normalization factor, and
\begin{equation}
\Omega_{\bf k}\equiv -{i\over 2}{f_k^\prime\over f_k}
\label{eqn:Omega}
\end{equation}
is written in terms of the function $f_k$, the solution to Equation~(\ref{eqn:uk}), i.e.,
\begin{equation}
f_k={1\over \sqrt{2k}}\left(1-{i\over k\tau}\right)e^{-ik\tau}\;.
\label{eqn:fk}
\end{equation}

The functional $\Psi[u(\tau,{\bf x})]$, representing a quantum
fluctuation in the inflaton field, is a superposition of many separate Fourier states
$\Psi_{\bf k}$ (Eqn.~\ref{eqn:Psitotal}). Somehow, this quantum description, which includes
interference between the various modes and no {\it a priori} classical correlations in phase
space between the field amplitudes and their canonical partners, must have transitioned into
classical fluctuations, characterized in part by a diagonalized density matrix and well defined
dynamical trajectories. How this might have happened is the issue we shall address next.

\subsection{The Quantum to Classical Transition}
The question concerning whether or not the state $\Psi[u(\tau,{\bf x})]$ can complete
the quantum to classical transition based on the factors we have summarized above has
received considerable attention over the past four decades, and is still an 
open one \cite{Halliwell:1985,Zurek:1992,Brandenberger:1992,Laflamme:1993,Polarski:1996,Grishchuk:1997,Hartle:1998,Kiefer:2000,Castagnino:2003,delCampo:2004,Martin:2005,Perez:2006}. As noted
in the introduction, the cosmological scalar perturbations must satisfy at least two
conditions to classicalize: (i) the system must undergo decoherence, so that quantum
interference becomes negligible and macroscopic variables appear. We see a universe with
well-defined, classical perturbation field values, not superpositions of multiple components
represented by Equation~(\ref{eqn:Psitotal}). (ii) One must see the establishment of classical
correlations in phase space, meaning that the allowed dispersion in the canonical variables
$(u,\pi)$ becomes insignificantly smaller than their classical orbital values.

Different authors have proposed somewhat different schemes for this transition, indicating
that no consensus has yet been reached. In part, this is due to the fact that the conventional
rules one may rely on in ordinary quantum mechanical applications are not necessarily all
available in cosmology. To begin with, the system being studied quantum mechanically is the
entire Universe. There is no possible separation into a subsystem of interest, its environment
and the observer. Moreover, there is just one universe, obviating any possibility of adopting
the statistical ensemble interpretation of quantum mechanical measurements. Even more importantly,
the observer is here a consequence of the quantum to classical transition, and could not
have played any causal part in it. For these (and other) reasons, the system being studied is
thus unusual by typical quantum mechanical standards.

A common step taken by the various approaches to solving this problem is to relegate most
of the very large number of degrees of freedom in $\Psi[u(\tau,{\bf x})]$ to an `environment'
and then to ignore them, allowing one to evolve the reduced density matrix $\hat{\rho}$ for the
subset of remaining (presumably more interesting) observables. Under some circumstances, also
involving suitable time averaging, this procedure eventually diagonalizes $\hat{\rho}$,
representing a complete mitigation of any interference effects, which is interpreted as the
emergence of classical behavior.

This is not entirely satisfactory, however, because decoherence has not yet successfully
solved the long-standing `measurement' problem in quantum mechanics. Certainly, the diagonalization
of $\hat{\rho}$ removes certain quantum traits from the system, but it is not clear that a
non-interfering set of simultaneous co-existing possibilities is necessarily classical
(see, e.g., ref.~\cite{Grib:1999}). Note, for example, that changing the Hilbert space of the
quantum system to a different basis would destroy the diagonal nature of the density matrix.
The interpretation of $\hat{\rho}$ is therefore subject to various observer-dependent choices.
What is lacking is a clear understanding of how to choose the basis and an interaction specific
eigenstate for the preferred observable. In our everyday experience, some progress may be made
by allowing the measurement device to `select' the basis, and adopting the ensemble interpretation
for the density matrix. But these features are obviously missing in the cosmological
context \cite{Hartle:1998,Hartle:2006}.

To solve the classicalization problem using decoherence in a cosmological setting, one must
therefore identify a physical mechanism and a preferential basis it selects.  It is also
necessary to find a criterion for separating the large number of degrees of freedom into
the `interesting set' and `the environment' dictated by the physical problem at hand.
At least some of the proposed treatments then appeal to `a specific realization' of the
stochastic variables \cite{Polarski:1996,delCampo:2004}, sounding very much like a
conventional `collapse' from the statistical description of the universe to one of the
members in the statistical ensemble. But no insight is provided into how and when such
a transition occurred in the real universe.

To address the second requirement, we begin by noting that in the remote conformal past, where
$\Omega_{\bf k}\rightarrow k/2$, the wavefunctional $\Psi_{\bf k}$ represented the ground state
of a harmonic oscillator, consistent with the previously described Bunch-Davies vacuum. For the
largest modes (those of relevance to the structure we observe today), the function $\Omega_{\bf k}$
acquired a non-trivial time dependence and Equation~(\ref{eqn:Psi}) evolved into a squeezed
state---meaning that, for these modes, there exists  a direction in the
$(u_{\bf k},\pi_{\bf k})$ plane where the dispersion is exponentially small, while the
dispersion in a perpendicular direction is very large. The resultant linear combination
of $u_{\bf k}$ and $\pi_{\bf k}$ for which the dispersion is minimized is often viewed
as the emerging classical phase-space trajectory.

A convenient tool to study this process, and its possible relevance to the classicalization
question, is the Wigner function (see, e.g., the reviews in 
refs.~\cite{Hillery:1984,Lee:1995}; for a more pedagogical account, see also 
ref.~\cite{Case:2008}), defined by
\begin{eqnarray}
{\mathcal{W}}(u_{\bf k}^R,u_{\bf k}^I,\pi_{\bf k}^R,\pi_{\bf k}^I)&=&{1\over (2\pi)^2}
\int dx\,dy\nonumber\\
&\null&\hskip-0.8in\Psi_{\bf k}\left(u_{\bf k}^R-{x\over 2},u_{\bf k}^I-{y\over 2}\right)
e^{-i\pi_{\bf k}^Rx-i\pi_{\bf k}^Iy}\times\qquad\nonumber\\
&\null&\hskip-0.4in\Psi_{\bf k}\left(u_{\bf k}^R+{x\over 2},u_{\bf k}^I+{y\over 2}\right)\;,
\label{eqn:Wigner}
\end{eqnarray}
in terms of the real and imaginary parts of $u_{\bf k}$ and $\pi_{\bf k}$.
Quantum mechanics inherently deals with probabilities, while classical physics deals with
well-defined trajectories in phase space. The Wigner function provides a means of representing
the density distributions in $u_{\bf k}$ and $\pi_{\bf k}$ for comparison with the ensemble
of trajectories one would get using classical means.

The literature on Wigner functions is extensive (including the aforementiond reviews
in refs.~\cite{Hillery:1984,Lee:1995,Case:2008}). Its properties suggest that it behaves like
a probability distribution in $(u,\pi)$ phase space, except that it can sometimes
take on negative values, so it is generally not a true probability distribution.
Moreover, points in the $(u,\pi)$ space to not represent actual states of the system
because the values of $u_{\bf k}$ and $\pi_{\bf k}$ cannot be determined precisely
at the same time. Nevertheless, ${\mathcal{W}}$ can be used to visualize correlations
between $u_{\bf k}$ and $\pi_{\bf k}$, particularly for Gaussian states, such as
we have in Equation~(\ref{eqn:Psi}), for which the Wigner function is indeed always
positive definite.

It is trivial to show in the case of Equation~(\ref{eqn:Psi}) that the explicit
form of ${\mathcal{W}}$ is
\begin{eqnarray}
{\mathcal{W}}(u_{\bf k}^R,u_{\bf k}^I,\pi_{\bf k}^R,\pi_{\bf k}^I)&=&{1\over \pi^2}
e^{-{\rm Re}\{\Omega_{\bf k}\}\left[(u_{\bf k}^R)^2+(u_{\bf k}^I)^2\right]}\times\nonumber\\
&\null&\hskip-1.3in
e^{-(\pi_{\bf k}^R+{\rm Im}\{\Omega_{\bf k}\}u_{\bf k}^R)^2/{\rm Re}\{\Omega_{\bf k}\}}
e^{-(\pi_{\bf k}^I+{\rm Im}\{\Omega_{\bf k}\}u_{\bf k}^I)^2/{\rm Re}\{\Omega_{\bf k}\}},\qquad
\label{eqn:squeeze}
\end{eqnarray}
with which one may clearly see the effect of strong squeezing as $\Omega_{\bf k}$ changes
dramatically with time (Eqns.~\ref{eqn:Omega} and \ref{eqn:fk}) during the cosmic expansion.

As long as $\Omega_{\bf k}\rightarrow k/2$, ${\mathcal{W}}$ is peaked over a small region
of phase space, representing the Wigner function of a coherent state, i.e., the ground state
of a harmonic oscillator. We shall return to this in \S~4 below, where we consider the corresponding
Wigner function for the numen quantum fluctuations. We shall see that this limiting situation takes on
added significance in that case, given that $\Omega_{\bf k}$ actually remains constant for a numen
field. But $\Omega_{\bf k}$ is definitely not constant here. As $\tau$ advances, ${\mathcal{W}}$
spreads and acquires a cigar shape typical of squeezed states \cite{Grishchuk:1990}, with a
drastically reduced dispersion around $\pi_{\bf k}$ and a correspondingly enormous dispersion
around $u_{\bf k}$. But the overall dispersion may be minimized, as we alluded to earlier, by
choosing an appropriate linear combination of $u_{\bf k}$ and $\pi_{\bf k}$, an outcome that
some argue represents the emerging correlation in classical phase space.

One may therefore think of a squeezed state as a state with the minimal uncertainty, though
not in terms of the original $(u,\pi)$ variables \cite{Polarski:1996,Kiefer:2000}.
In the cosmological context, the inflationary expansion created an uncertainty on the
value of the field and its conjugate momentum that was much larger than the minimum
uncertainty implied by the Heisenberg uncertainty principle. The minimum uncertainty was
instead associated with a new pair of `rotated' canonical variables. But this outcome
has also been challenged as not necessarily representing a classical system
\cite{Laflamme:1993,Perez:2006}.

Take the following situation as an example. Consider an electron in a minimal wavepacket
localized at the origin, with an uncertainty $\Delta x$ in position and
$\Delta p=\hbar/(2\Delta x)$ in momentum. Then form a superposition of this state with an
identical one after a translation by a large distance $D$. The overall uncertainty is
now $\sim D\hbar/(2\Delta x)$, which can be increased arbitrarily by simply choosing a
sufficiently large distance $D$. This situation is clearly analogous to one of our
squeezed states, but the superposition we have created is nonetheless not classical.

And then there is the issue of definite outcomes, also known as the `measurement
problem' \cite{Colanero:2012}, as we alluded to in the introduction. This aspect of quantum 
mechanics has been with us for over a century. Even if decoherence were successful in 
diagonalizing the density matrix, it cannot solve the definite outcome problem,
which is far worse in cosmology than it is in typical laboratory situations.
The usual approach of adopting the Copenhagen interpretation of quantum mechanics,
in which a measurement `collapses' the state vector of the system into an eigenstate
corresponding to the measurement result, does not work in the early Universe, where
there were no measurement devices or observers present.

While decoherence might have diagonalized the reduced density matrix of the system
to an ensemble of classically observable universes, it does not explain how a
certain universe was singled out to be observed \cite{Sudarsky:2011}. Perhaps the
Copenhagen interpretation is just not well suited to cosmology. In addition, if
decoherence resulted from an environment comprised of the `non-interesting' degrees
of freedom in $\Psi[u(\tau,{\bf x})]$, how could this happen when inflation would
have driven all such fields towards their vacuum states? Moreover, why do we have
the privilege of deciding which degrees of freedom to relegate to the background
based solely on whether our current technology allows us to observe them today?

We shall now divert our attention away from the traditional inflationary scenario
we have been describing, and seriously consider the implications of a novel feature
emerging from the latest release of the {\it Planck} data: the observational
measurement of a cutoff in the primordial spectrum bears directly on the question
of how and when quantum fluctuations were generated in the early Universe. We shall
study what new ideas and constraints this brings to the classicalization process.

\section{Quantum Fluctuations at the Planck Scale}
\subsection{Emergence of a Cutoff in the primordial power spectrum ${\mathcal{P}}(k)$}
All three of the major satellite missions designed to study the CMB---COBE
\cite{Hinshaw:1996}; WMAP \cite{Bennett:2003}; and {\it Planck}
\cite{Planck:2018}--- have uncovered several anomalies in its fluctuation spectrum. The
two most prominent among them are: (1) an unexpectedly low level---perhaps even a
complete absence---of correlation at large angles (i.e., $\theta\gtrsim 60^\circ$),
manifested via the angular correlation function, $C(\theta)$; and (2) relatively
weak power in the lowest multipole moments of the angular power spectrum, $C_\ell$.
Their origin, however, is still subject to considerable debate, many arguing in favor
of a misinterpretation or the result of unknown systematics, such as an incorrect
foreground subtraction (see refs.~\cite{Bennett:2011,Copi:2010} for reviews). This
uncertainty is also fueled in large part by a persistent lack of clarity concerning
how quantum fluctuations were seeded in the early universe \cite{Melia:2019a,LiuMelia:2020}.

The large-scale anomalies stand in sharp contrast to our overall success
interpreting the CMB anisotropies at angles smaller than $1^\circ$. Over
the past several decades, a concerted effort has therefore been made in attempting to identify
features in the primordial power spectrum, ${\mathcal{P}}(k)$, responsible for their origin.
For example, in their study of the CMB angular power spectrum, Shafieloo \& Souradeep
\cite{Shafieloo:2004} assumed an exponential cutoff at low $k$-modes, and identified a 
turnover generally consistent with the most recent measurement we shall discuss below 
(see Eq.~\ref{eqn:kmin}). This early treatment was based on WMAP observations 
\cite{Bennett:2003}, however, not the higher precision {\it Planck} measurements 
\cite{Planck:2018} we have today, so the reality of a non-zero $k_{\rm min}$ remained 
somewhat controversial.

In followup work, Nicholson et al. \cite{Nicholson:2009} and Hazra et al. \cite{Hazra:2014} 
inferred a `dip' in ${\mathcal{P}}(k)$ on a scale $k\sim 0.002$ Mpc$^{-1}$ for the WMAP data,
confirming the outcome of an alternative approach by Ichiki et al. \cite{Ichiki:2010} that 
identified an oscillatory modulation around $k\sim 0.009$ Mpc$^{-1}$. In closely aligned work, 
Tocchini et al. \cite{Tocchini:2005} modeled both a dip at $k\sim 0.035$ Mpc$^{-1}$ and a `bump' 
at $k\sim 0.05$ Mpc$^{-1}$. Like the others, though, these features were based solely on WMAP 
observations and therefore appeared to be merely suggestive rather than compelling. Tocchini 
et al. \cite{Tocchini:2006} improved on this analysis considerably, and found evidence for 
three features in ${\mathcal{P}}(k)$, one of which was a cutoff at $\sim 0.0001-0.001$ 
Mpc$^{-1}$ at a confidence level of $\sim 2\sigma$. Complementary work by Hunt \& Subir
\cite{Hunt:2014,Hunt:2015} confirmed the likely existence of a cutoff $k < 5 \times 
10^{-4}$ Mpc$^{-1}$ but, as before, also concluded that more accurate {\it Planck} 
data would eventually be needed to confirm these results more robustly. 

The subsequent {\it Planck} observations have not only largely confirmed these earlier
results, but have provided us with a greatly improved precision in the `measurement'
of $k_{\rm min}$. For example, the Planck Collaboration \cite{Planck:2016} fit a cutoff 
to the CMB angular power spectrum and found a value $\sim 3-4\times 10^{-4}$ Mpc$^{-1}$.
Still, though these studies all pointed to the likely existence of a cutoff in 
${\mathcal{P}}(k)$, a non-zero $k_{\rm min}$ could not be claimed with a confidence
level exceeding $1-2\sigma$.

This situation improved considerably when, instead of looking solely at the power spectrum,
the impact of a cutoff was also considered on the angular correlation function, independently
of the angular power spectrum. Two separate 
(though complementary) studies of the latest {\it Planck} data release have provided more
compelling evidence that the two large-angle features in the CMB anisotropies may be real. 
The first of these \cite{MeliaLopez:2018} demonstrates that the most likely explanation for 
the missing large-angle correlations is a cutoff,
\begin{equation}
k_{\rm min}={4.34\pm0.5\over r_{\rm cmb}}\;,\label{eqn:kmin}
\end{equation}
in the primordial power spectrum, ${\mathcal{P}}(k)$, where $r_{\rm cmb}$
is the comoving distance to the surface of last scattering. For the {\it Planck}-$\Lambda$CDM
parameters, $r_{\rm cmb}\approx 13,804$ Mpc \cite{Planck:2018}, and we therefore have
$k_{\rm min}=(3.14\pm 0.36)\times 10^{-4}$~Mpc$^{-1}$. A zero cutoff (i.e., $k_{\rm min}=0$)
is ruled out by the $C(\theta)$ data at a confidence level exceeding $\sim 8\sigma$.

A subsequent study \cite{Melia:2020a} focused on the CMB angular power spectrum itself
(i.e., $C_{\ell}$ versus $\ell$), and its results (i) confirmed that the introduction of
this cutoff in ${\mathcal{P}}(k)$ does not at all affect the remarkable consistency
between the standard inflationary model prediction and the {\it Planck} measurements
at $\ell>30$ \cite{Planck:2018}, where the underlying theory is widely believed to be
correct; and (ii) showed that such a cutoff ($k_{\rm min}$) also self-consistently explains
the missing power at large angles, i.e., the low multipole moments ($\ell=2-5$). The cutoff
optimized by fitting the angular power spectrum over the whole range of $\ell$'s is
$k_{\rm min}=(2.04^{+1.4}_{-0.79}) \times 10^{-4}$~Mpc$^{-1}$, while a fit to the restricted
range $\ell\le 30$, where the Sachs-Wolfe effect is dominant \cite{Sachs:1967}, gives
$k_{\rm min}=(3.3^{+1.7}_{-1.3})\times 10^{-4}$~Mpc$^{-1}$. The outcome based on the CMB
angular power spectrum therefore rules out a zero cutoff at a confidence level
$\gtrsim 2.6\sigma$. In either case, the inferred value of $k_{\rm min}$ is fully
consistent with the cutoff implied by missing correlations in $C(\theta)$, and one
concludes that both of these large-angle anomalies are probably due to the same truncation,
i.e., $k_{\rm min}\sim 3\times 10^{-4}$ Mpc$^{-1}$, in ${\mathcal{P}}(k)$. The
confidence with which one may make such a claim depends on whether the cutoff is used to
address the missing power at low $\ell$'s, or the missing correlations at large
angles. Nevertheless, the fact that the same $k_{\rm min}$ apparently solves both
anomalies makes the assumption of a cutoff quite reasonable.

These results reinforce the perception that the small-angle anisotropies (for $\ell\gtrsim 30$),
which are mostly due to acoustic oscillations, are well understood, while the fluctuations
associated with angular correlations at $\theta\gtrsim 60^\circ$, due to the
Sachs-Wolfe effect, continue to be problematic for the standard inflationary picture
\cite{Hinshaw:1996,Bennett:2003,Planck:2018}. The evidence for a non-zero $k_{\rm min}$
speaks directly to the cosmological expansion itself. At $\ell\lesssim 30$, we are probing ever
closer to the beginning of inflation, culminating with the cutoff $k_{\rm min}$, signaling
the very first mode that would have crossed the horizon when the quasi-de Sitter 
phase started \cite{Mukhanov:1992,LiuMelia:2020,Melia:2020b}.

But if one insists on inflation simultaneously fixing the horizon problem and accounting
for the observed primordial power spectrum, ${\mathcal{P}}(k)$, the implied time at which
the accelerated expansion began would have suppressed the comoving size of the universe
to a tenth of the required value \cite{LiuMelia:2020,Melia:2020b}. Moreover,
neither a radiation-dominated, nor a kinetic-dominated, phase preceding inflation could have
alleviated this disparity \cite{Destri:2008,Destri:2010}.

\subsection{A Reinterpretation of $k_{\rm min}$}
To be clear, the measurement of $k_{\rm min}$ does not completely rule out inflation, nor even the
idea that some slow-roll variant may eventually be constructed to address the inconsistency described
above. At a minimum, however, the currently proposed inflaton potentials $V(\phi)$ require at
least some modification. Moreover, a cutoff in ${\mathcal{P}}(k)$ does not argue {\it against}
the influence of a scalar field, $\phi$, nor anisotropies arising from its quantum fluctuations,
but there is now some motivation to question whether $V(\phi)$ was truly inflationary.

Over the past decade, some evidence has been accumulating that the cosmic fluid may possess
a zero active mass equation-of-state, $\rho+3p=0$ (in terms of its total energy density
$\rho$ and pressure $p$), supported by over 27 different kinds of observation at low and high
redshifts (see Table~2 in ref.~\cite{Melia:2018a} for a recent summary of these results).
Such a universe lacks a horizon problem \cite{Melia:2013,Melia:2018b}, so the lack of a fully
self-consistent inflationary paradigm may be telling us that the universe does not need it.
This is the key assumption we shall make to reinterpret $k_{\rm min}$ in this paper. In
addition to the growing body of empirical evidence favoring this approach, there is also
theoretical support for the zero active mass equation-of-state from the `Local Flatness
Theorem' in general relativity \cite{Melia:2019b}.

As noted earlier, we shall clearly distinguish between the roles played by a non-inflaton 
$\phi$ and a conventional inflaton field by informally refering to the former as a `numen' 
field, based on our inference that it may represent the earliest form of substance in the 
universe. Its equation-of-state is assumed to be $\rho_\phi+3p_\phi=0$, and we shall see 
shortly why this property appears to provide a more satisfactory interpretation of 
$k_{\rm min}$ than an inflaton field.

The background numen field is homogeneous, so its energy density $\rho_\phi$ and
pressure $p_\phi$ are simply given as
\begin{equation}
\rho_\phi={1\over 2}{\dot{\phi}}^2+V(\phi)\;,
\end{equation}
and
\begin{equation}
p_\phi={1\over 2}{\dot{\phi}}^2-V(\phi)\;.
\end{equation}
The zero active mass equation-of-state therefore uniquely constrains the potential to be
\begin{equation}
V(\phi)={{\dot{\phi}}^2}\;,
\label{eqn:potnumen}
\end{equation}
with the explicit solution
\begin{equation}
V(\phi)=V_0\,\exp\left\{-{2\sqrt{4\pi}\over m_{\rm Pl}}\,\phi\right\}\;,
\label{eqn:numenpot}
\end{equation}
in terms of the Planck mass
\begin{equation}
m_{\rm Pl}\equiv {1\over\sqrt{G}}\;.\label{eq:Planckmass}
\end{equation}
Some may recognize this as a special member of the category of minimally coupled fields
explored in the 1980's \cite{Abbott:1984,Lucchin:1985,Barrow:1987,Liddle:1989}, intended
to produce so-called power-law inflation. But unlike the other fields in this cohort, the
numen field's zero active mass equation-of-state makes it the only member of this group
that does {\it not} inflate, since the Friedmann equations with this density and pressure
lead to an expansion factor $a(t)=t/t_0$, written in terms of the age of the universe,
$t_0$. This normalization of $a(t)$ is appropriate for a spatially flat FLRW metric,
which the observations appear to be telling us.

With this expansion factor, the conformal time may be written
\begin{equation}
\tau(t)=t_0\ln a(t)\;, 
\label{eqn:tau}
\end{equation}
such that the zero of $\tau$ coincides with $t=t_0$. The parameter $z$ in Equation~(\ref{eqn:z})
thus becomes
\begin{equation}
z={m_{\rm Pl}\over \sqrt{4\pi}}a(t)\;,
\end{equation}
so that $z^\prime/z=1/t_0$ and $z^{\prime\prime}/z=1/t_0^2$.
The resulting curvature perturbation equation analogous to Equation~(\ref{eqn:uk})
may thus be written
\begin{equation}
u_{\rm k}^{\prime\prime}+\alpha_{\rm k}^2 u_{\rm k}=0\;,
\label{eqn:uknumen}
\end{equation}
where the frequency (analogous to Eqn.~\ref{eqn:freq}) is now given by the expression
\begin{equation}
\alpha_{\rm k}\equiv {1\over t_0}\sqrt{\left(2\pi R_{\rm h}\over \lambda_{\rm k}\right)^2-1}\;,
\label{eqn:freqnumen}
\end{equation}
in terms of the proper wavelength of mode $k$,
\begin{equation}
\lambda_{\rm k}\equiv {2\pi\over k} a(t)\;.
\end{equation}
In this expression, the quantity $R_{\rm h}\equiv c/H=ct$ is the apparent (or
gravitational) radius \cite{Melia:2018c}, which defines the Hubble horizon in
a spatially flat universe. The most critical difference between Equations~(\ref{eqn:uk})
and (\ref{eqn:freq}), and Equations~(\ref{eqn:uknumen}) and (\ref{eqn:freqnumen}), is
that here both $R_{\rm h}$ and $a(t)$ scale linearly with $t$, and therefore the frequency
$\alpha_{\rm k}$ of the numen quantum fluctuations is {\it always time-independent}.
This is because the ratio $R_{\rm h}/\lambda_k\propto kR_{\rm h}/a(t)$ is constant
for each $k$, and therefore numen fluctuations do not criss-cross the horizon; once
$\lambda_{\rm k}$ is established upon the mode's exit into the semi-classical universe,
it remains a fixed fraction of $R_{\rm h}$ as they expand with time. This feature is
crucial to understanding how and why numen quantum fluctuations provide a more
satisfactory explanation than inflation for the origin of $k_{\rm min}$.

The solution to Equation~(\ref{eqn:uknumen}) is that of the standard harmonic
oscillator:
\begin{equation}
u_{\rm k}(\tau) = \left\{ \begin{array}{ll}
         B(k)\,e^{\pm i\alpha_{\rm k}\tau} & \mbox{($2\pi R_{\rm h}>\lambda_{\rm k}$)} \\
         B(k)\,e^{\pm |\alpha_{\rm k}|\tau} & \mbox{($2\pi R_{\rm h}<\lambda_{\rm k}$)}\end{array} \right. \;.
\label{eqn:numenmode}
\end{equation}
That is, all modes with $\lambda_{\rm k}<2\pi R_{\rm h}$ oscillate, while the super-horizon
ones do not, mirroring the behavior of the more conventional inflaton field. Here, however,
the mode with the longest wavelength relevant to the formation of structure is
the one for which $\lambda_{\rm k}(\tau)=2\pi R_{\rm h}(\tau)$, i.e.,
\begin{equation}
k^{\rm numen}|_{\rm min}=1/t_0\;.
\end{equation}
One's intuition would immediately suggest that $k^{\rm numen}|_{\rm min}$ ought to be identified
with the cutoff $k_{\rm min}$ measured in the CMB, and it is not difficult to demonstrate why
that has to be the case for a numen scalar field, as we shall see shortly.

A slightly different (and simpler) definition of the Planck scale than that appearing in 
Equation~(\ref{eq:Planckmass}) is based on the length $\lambda_{\rm Pl}$ at which the Compton 
wavelength $\lambda_{\rm C}\equiv 2\pi/m$ for mass $m$ equals its Schwarzschild radius 
$R_{\rm h}\equiv 2Gm$. That is,
\begin{equation}
\lambda_{\rm Pl}\equiv \sqrt{4\pi G}\;.
\end{equation}
The Compton wavelength $\lambda_{\rm C}$ grows as the gravitational radius $R_{\rm h}$ shrinks,
so the standard inflationary picture conflicts with quantum mechanics in its interpretation of
wavelengths shorter than $2\pi\lambda_{\rm Pl}$ (the factor $2\pi$ arising from the definition
of $\lambda_{\rm Pl}$ in terms of $R_{\rm  h}$). This is a serious problem for the standard
model because the fluctuation amplitude $A_s$ measured in the CMB anisotropies requires
quantum fluctuations in the inflaton field to have been born in the Bunch-Davies vacuum,
long before the Planck time $t_{\rm Pl}\equiv\lambda_{\rm Pl}$, a conundrum commonly referred
to as the ``Trans-Planckian Problem" \cite{Martin:2001}.

The numen field can completely avoid this inconsistency if we argue that each mode $k$ emerged
into the semi-classical universe when
\begin{equation}
\lambda_{\rm k}=2\pi\lambda_{\rm Pl}\;,
\end{equation}
and then evolved subject to the oscillatory solution in Equation~(\ref{eqn:numenmode}). As we shall
see shortly, the fact that each succeeding $k$-mode emerges at later times in this picture produces 
a near scale-free power spectrum ${\mathcal{P}}(k)$ with $n_s\lesssim 1$ \cite{Melia:2019a}. Such 
an idea---that modes could have been born at a particular spatial scale---has already received some 
attention in the past, notably by Hollands \& Wald \cite{Hollands:2002}. In their case, however, the 
fundamental scale was not related to $\lambda_{\rm Pl}$. Others supporting this proposal include
Brandenberger et al. \cite{Brandenberger:2002} and Hassan et al. \cite{Hassan:2003}.

It is not difficult to understand why the fundamental scale for the numen field must be
$\lambda_{\rm Pl}$. If we interpret $k^{\rm numen}|_{\rm min}$ to be $k_{\rm min}$,
the latter defines the time $t_{\rm min}$ at which the first quantum fluctuation emerged
out of the Planck domain into the semi-classical universe.  From Equation~(\ref{eqn:kmin})
and the expression for $r_{\rm cmb}$ in a universe with zero active mass,
\begin{equation}
r^{\rm numen}|_{\rm cmb}={c\over H_0}\ln(1+z_{\rm cmb})\;,
\end{equation}
one therefore finds that
\begin{equation}
t_{\rm min}={4.34\,t_{\rm Pl}\over \ln(1+z_{\rm cmb})}\;.
\end{equation}
Its dependence on $z_{\rm cmb}$ is so weak that $t_{\rm min}$ is approximately
equal to $t_{\rm Pl}$ regardless of where the last scattering surface was located.
For example, in {\it Planck}-$\Lambda$CDM, $z_{\rm cmb}\sim 1080$, for which
$t_{\rm min}\sim 0.63\, t_{\rm Pl}$. But even if we were to adopt a very different value
$z_{\rm cmb}=50$, the first quantum fluctuation would have emerged at
$t_{\rm min}\sim 1.1\, t_{\rm Pl}$.

If the spatially largest fluctuation we see in the CMB was due to a numen fluctuation,
one concludes from this analysis that it must have emerged out of the Planck regime
at roughly the Planck time. In other words, this fluctuation would have physically
exited into the semi-classical universe shortly after the Big Bang---indeed, it would
have appeared as soon as it could, given what we currently understand about the Planck
time $t_{\rm Pl}$. No other scalar field introduced thus far, inflaton or otherwise, has
this very interesting property.

But if a Bunch-Davies vaccum in the remote conformal past is not used
for these fluctuations, how does one then determine the normalization constant $B(k)$
of the modes in Equation~(\ref{eqn:numenmode})? As we saw in \S~2.1, a principal complication
with the inflaton field is the significant spacetime curvature encountered by its quantum
fluctuations as they cross the Hubble radius. This led to the introduction of a Bunch-Davies
vacuum in the distant conformal past, where the modes could have been seeded in
Minkowski space. We made reference to the fact that this situation actually creates
an inconsistency with quantum mechanics, often referred to as the trans-Planckian problem.
This issue is largely beyond the scope of the present paper, however, because the numen field
completely avoids this inconsistency.

Even though the numen quantum fluctuations emerged at the Planck scale---with a wavelength
comparable to the gravitational radius $R_{\rm h}$---the zero active mass equation-of-state
in the cosmic fluid (leading to Eqn.~\ref{eqn:potnumen}) ensures that the frame into which they
emerged from the Planck regime was geodesic. That is, in spite of the Hubble expansion, the universe
was always in free fall, with {\it zero internal acceleration}. One can easily confirm this
from the fact that the frequency $\alpha_{\rm k}$ in Equation~(\ref{eqn:freqnumen}) is always
time-independent. One therefore does not need an ad hoc construction of a Bunch-Davies
vacuum, and we may simply set
\begin{equation}
B(k)={1\over \sqrt{2\alpha_{\rm k}}}
\label{eqn:numennorm}
\end{equation}
for the numen quantum fluctuations (in Eqn.~\ref{eqn:numenmode}), following the same
minimization of the Hamiltonian argument that led to Equation~(\ref{eqn:Mink}).

We complete this brief survey by considering the spectrum ${\mathcal{P}}(k)$ one 
should expect from the birth of quantum fluctuations at the Planck scale. Right away, we can 
see from Equation~(\ref{eqn:freqnumen}) that the difference between $\alpha_{\rm k}$ and $k$ 
is what distorts the primordial power spectrum away from a pure, scale-free distribution, as 
we shall confirm below. This is most easily recognized if we rewrite the mode frequency in the
form $\alpha_{\rm k}=\sqrt{k^2-{1/t_0^2}}$ (which is $<k$).

From the definition of $u_{\rm k}$ and the curvature perturbation, we see that
\begin{equation}
|{\mathcal{R}}_{\rm k}|^2={2\pi\over m_{\rm Pl}^2}{1\over \alpha_{\rm k} a^2}\;.
\end{equation}
The power spectrum ${\mathcal{P}}_{\mathcal{R}}(k)$ is defined as 
\begin{equation}
{\mathcal{P}}_{\mathcal{R}}(k)\equiv {k^3\over 2\pi^2}|{\mathcal{R}}_{\rm k}|^2\;,
\end{equation}
and therefore 
\begin{equation}
{\mathcal{P}}_{\mathcal{R}}(k)={1\over (2\pi)^2}\left[{a(t_k)\over a(t)}\right]^2\left[1-
\left({k_{\rm min}\over k}\right)^2\right]^{-1/2}\;.\label{eqn:spec1}
\end{equation}
What is not known yet is how these quantum fluctuations devolved into grand unified theories 
particles, after which the perturbation amplitude remained frozen. Nevertheless, it is likely that 
the dynamics of this decay/evolution is associated with a particular length (or energy) scale, 
$L_*$ \cite{Brouzakis:2012,Dvali:2012}, not unlike the Planck scale $\lambda_{\rm Pl}$.

Mode $k$ reached this scale at $a(t_k^*)=L_*k/2\pi$ or, equivalently, at
time $t_k^*=t_0L_*k$. And so Equation~(\ref{eqn:spec1}) may be re-written
\begin{equation}
{\mathcal{P}}_{\mathcal{R}}(k)={1\over (2\pi)^2}\left[{\lambda_{\rm Pl}\over L_*}\right]^2\left[1-
\left({k_{\rm min}\over k}\right)^2\right]^{-1/2}\;,\label{eqn:spec2}
\end{equation}
which one then needs to compare with the observed CMB power spectrum ${\mathcal{P}}(k)=A_s(k/k_0)^{n_s-1}$. 
In the context of the standard model, the {\it Planck} optimizations \cite{Planck:2018} give
$A_s=(2.1\pm 0.04) \times 10^{-9}$ and $n_s=0.9649\pm 0.0042$. And it is not difficult to 
see that $L_*$$\sim$$3.5\times 10^3 \lambda_{\rm Pl}$. Therefore, with the Planck scale set at 
$m_{\rm Pl}\approx 1.22\times 10^{19}$ GeV, one finds that $L_*$ corresponds to an energy
of roughly $3.5\times 10^{15}$ GeV, remarkably consistent with the energy 
scale expected in grand unified theories. Of course, much of this is mere speculation at
the present time, given that the physics of this process lies beyond the standard model.
Nevertheless, the quantum fluctuations in this picture would have oscillated until 
$t\sim 3.5\times 10^3t_{\rm Pl}$, after which the numen field would have devolved into 
grand unified theory particles, with a freezing of the perturbation amplitude thereafter.

The primordial power spectrum (Eq.~\ref{eqn:spec2}) is almost scale-free, but
not exactly. Using the conventional definition of the scalar index, we find that 
\begin{equation}
n_s=1+{d\,\ln {\mathcal{P}}_{\mathcal{R}}(k)\over d\,\ln k}= 1-{1\over (k/k_{\rm min})^2-1}\;.\label{eqn:ns}
\end{equation}
The index $n_s$ is therefore slightly less than $1$, and we confirm that the deviation from 
a pure scale-free distribution is due to the aforementioned difference between $\alpha_{\rm k}$ 
and $k$, which in the end arises from the Hubble expansion (or `frictional') term 
${\mathcal{R}}^\prime_k$ in the growth Equation~(\ref{eqn:mathcalR}).

At least qualitatively, this result agrees with the value of $n_s$ measured by {\it Planck}.
Of course, the numen optimization may produce a different outcome than that seen with
$\Lambda$CDM, but it would be difficult to see why the `red' tilt ($n_s<1$) should be converted
into `blue' ($n_s>1$) with a change in background cosmology. An average of $n_s$ over
$k$ in Equation~(\ref{eqn:ns}) gives $\langle n_s\rangle\sim 0.96$ over the range
$k_{\rm min}\lesssim k\lesssim 50k_{\rm min}$. The spectral index would approach closer to
one at larger values of $k$.

The caveat here, of course, is that there are still several unknowns with this process that
make it impossible for us to know all the factors influencing the value of $n_s$. For example,
we have assumed that the length scale at which the modes emerge from the Planck regime 
is always fixed at $\lambda_{\rm Pl}$. But this constraint depends critically on the nature 
of quantum gravity and its transition into general relativity. It may turn out that the actual
length scale varies as the Universe expands. If so, this evolution would produce an additional 
deviation of $n_s$ from one. As a quick illustration, suppose we were to represent such a 
variation with the scaling $\lambda_{\rm PL}\rightarrow \lambda_{\rm Pl}k^{-\beta}$. In that 
case, Equation~(\ref{eqn:ns}) would become
\begin{equation}
n_s= 1-2\beta-{1\over (k/k_{\rm min})^2-1}\;,\label{eqn:nsbeta}
\end{equation}
and the {\it Planck} data would then imply that $\beta\lesssim 0.02$.

\section{Classicalization of Quantum Fluctuations at the Planck Scale}
\subsection{The Birth of Quantum Fluctuations at the Planck Scale}
The acute classicalization problem plaguing the inflaton field stems directly from the nature
of the wave functional $\Psi[u(\tau,{\bf x})]$ describing its fluctuations (Eqn.~\ref{eqn:Psitotal}).
As discussed in \S~2.2, there is no consensus yet on how the interference between its mode
components could have been completely removed.  This problem does not exist for numen quantum
fluctuations due to the way they were seeded---unlike their inflaton counterparts, each of the
numen fluctuations emerged with a single $k$-mode. Pure, single $k$-mode quantum states were
established from the very beginning as a result of the distinct {\it time} at which they entered
the semi-classical universe out of the Planck domain.

A numen fluctuation {\bf b}orn at time $t_{\rm b}$, had a unique comoving wavenumber defined
by the relation
\begin{equation}
k_{\rm b}\equiv {2\pi a(\tau_{\rm b})\over \lambda_{\rm Pl}}
\end{equation}
or, more explicitly,
\begin{equation}
k_{\rm b}={2\pi\over \lambda_{\rm Pl}\,t_0}t_{\rm b}\;.
\label{eqn:kb}
\end{equation}
Thus, the oscillating modes in Equation~(\ref{eqn:numenmode}) should more accurately be
written as follows:
\begin{equation}
u_{\rm k}(\tau,\tau_{\rm b})={1\over\sqrt{2\alpha_{\rm k}}}e^{\pm i\alpha_{\rm k}\tau}\delta(k-k_{\rm b})\;,
\label{eqn:uknumen2}
\end{equation}
where clearly $k_{\rm b}$ is uniquely related to $\tau_{\rm b}$ via Equations~(\ref{eqn:tau})
and (\ref{eqn:kb}). In other words, it is not sufficient to merely track the temporal evolution
of a numen quantum fluctuation. One must also specify its time of birth.

For the numen quantum fluctuation operator, one thus has
\begin{eqnarray}
\hat{u}(\tau,\tau_{\rm b},{\bf x})&=&{1\over (2\pi)^{3/2}}\int {d^3{\bf k}\over \sqrt{2\alpha_{\rm k}}}
\delta(k-k_{\rm b})\times\nonumber\\
&\null&\left[\hat{a}_{\bf k}^- e^{i({\bf k}\cdot{\bf x}-\alpha_{\rm k}\tau)}+
\hat{a}_{\bf k}^+ e^{-i({\bf k}\cdot{\bf x}-\alpha_{\rm k}\tau)}\right].\qquad
\end{eqnarray}
The integral is straightforward to evaluate, and one finds that the numen field operator,
analogous to Equation~(\ref{eqn:fieldop}) for the inflaton field, is
\begin{equation}
\hat{u}(\tau,\tau_{\rm b},{\bf x})={k_{\rm b}^2\over \sqrt{\pi\,\alpha_{{\rm k}_{\rm b}}}}\,
j_0(k_{\rm b}|{\bf x}|)\left[\hat{a}_{{\bf k}_{\rm b}}^-e^{-i\alpha_{{\rm k}_{\rm b}}\tau}
+\hat{a}_{{\bf k}_{\rm b}}^+e^{i\alpha_{{\rm k}_{\rm b}}\tau}\right]\;,
\label{eqn:Bess}
\end{equation}
where $j_0(k_{\rm b}|{\bf x}|)$ is a spherical Bessel function of the first kind.
We emphasize again that, in this picture, the entire numen quantum fluctuation born
at $\tau_{\rm b}$ is characterized by a single wavenumber $k_{\rm b}$. A schematic
diagram showing the amplitude of this operator at any given time, $\tau\ge \tau_{\rm b}$,
is shown in figure~1.

\begin{figure}
\vskip 0.1in
\centering
\includegraphics[width=1.0\linewidth]{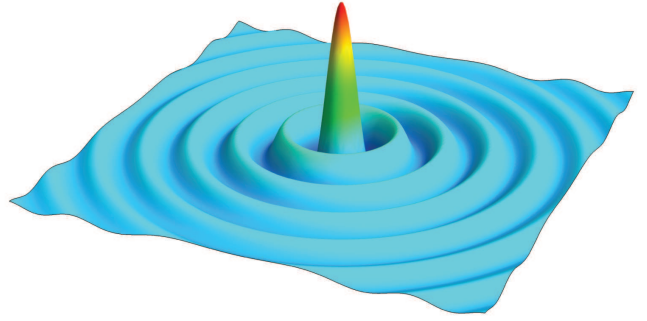}
\caption{Schematic diagram showing the amplitude (plotted along $\hat{z}$) of a numen field
operator projected onto the $xy$-plane. The spatial variation (in comoving coordinates {\bf x})
is proportional to the spherical Bessel function of the first kind, $j_0(k_{\rm b}|{\bf x}|)$,
where $k_b$ is the unique wavenumber corresponding to the time $\tau_{\rm b}$ at which the
fluctuation emerged into the semi-classical universe across the Planck scale $\lambda_{\rm Pl}$
(see Eqns.~\ref{eqn:kb} and \ref{eqn:Bess}).}
\end{figure}

Several of the characteristics we see in the numen field were anticipated by the model
proposed by Hollands and Wald \cite{Hollands:2002}, though their basic concept and scales
were quite different. The essence underlying the numen field's behavior is its inferred potential
given in Equation~(\ref{eqn:numenpot}). As in our case, Hollands and Wald also started by
considering the dynamics of a scalar field in a background cosmological setting, though the
equation-of-state in their cosmic fluid was inconsistent with the zero active mass
condition. Nevertheless, in both their and our cases we require that the fluctuations were
born in their ground state at a fixed length scale. The work described in this paper
is strongly informed by the recent {\it Planck} data release and its subsequent analysis,
which were not available at the time Hollands and Wald proposed their model. As we have
seen, the latest measurements strongly suggest that the cutoff in the observed primordial
spectrum is associated with the very first quantum fluctuation that exited the Planck
domain at about the Planck time. There is no room, in this proposal, for the basic length
scale at which the numen fluctuations were born to be anything but the Planck wavelength.

But though the proposal by Hollands and Wald lacked the rigor imposed by the latest
{\it Planck} observations, already Perez et al. \cite{Perez:2006} noted a very important
trait of quantum fluctuations born in this fashion. They pointed out that such a model
clearly demonstrates the need for some process to be responsible for their birth,
playing a role analogous to that of quantum mechanical {\it measurement}. The birth of
the mode, they argued, is effectively the step in which quantum mechanical uncertainty
is removed. This is precisely the situation we describe in Equation~(\ref{eqn:uknumen2}),
whereby each quantum fluctuation emerging into the semi-classical universe is
characterized by a unique wavenumber, constrained by the time at which the fluctuation
was created.

Perez et al. \cite{Perez:2006} also wondered what process could be associated
with the particular time of such an occurrence. With the insights we have gained from the
{\it Planck} data, we can now suggest that the length scale at which the numen modes
were born is not at all random. It carries significant physical meaning. As we have seen,
the Planck length is the scale at which the Compton wavelength equals the Schwarzschild
radius, meaning that below this scale is the realm of quantum gravity. If the picture
we are describing in this paper is correct, the birth of numen quantum fluctuations was
associated with the release of these modes into the semi-classical universe, where
gravity is adequately described by (the classical theory of) general relativity. Moreover,
as far as we know, the Planck scale never changes. Thus, as quantum fluctuations
stochastically exited in time, their wavenumber had to reflect their time of `birth'
and it is this correlation that built the near scale-free power spectrum seen in the
cosmic microwave background \cite{Planck:2018,Melia:2019a}.

The manner in which numen fluctuations were born thus already removes a major
hurdle in the classicalization process. Rather than having to deal with quantum mechanical
interference between many degrees of freedom, here we have distinct quantum fluctuations
possessing unique wavenumbers. Two issues remain, however, one having to do with how these
modes acquired classical correlations in phase space and the mechanism that converted a
homogeneous, isotropic universe into an inhomogeneous one. We shall consider the latter
next, and then revisit the Wigner function (Eqn.~\ref{eqn:Wigner}) to resolve the former.

\subsection{Anisotropies}
A missing ingredient from much of the past discussion concerning classicalization in the early
universe has been the process by which a perfectly homogeneous and isotropic state transformed
into an inhomogeneous and anisotropic state described by the density fluctuations. A chief
reason for this handicap has been the absence of an obvious `external' source of asymmetry.
Broadly speaking, such a transition can be effected as part of an `R' process, e.g.,
measurement or collapse, but not a U process, i.e., a unitary evolution via a Schr\"odinger
type of equation. It has been recognized that without a measurement-like process, the
required transition could not have happened. In hindsight, the problem has actually been
a lack of appreciation for the importance of the gravitational (or `apparent') horizon
$R_{\rm h}(\tau)=c/H(\tau)$ \cite{Melia:2018c}.

Yes, the universe is homogeneous and isotropic, but only when described using
the `community' coordinates of myriads of observers dispersed throughout the cosmos
\cite{Weinberg:1972}. But from the perspective of a single observer at a fixed spacetime
point, the universe does not appear to be homogeneous. His/her description of the physical
state of the system, using their coordinates centered at their location, must take into account
the effects of spacetime curvature. Take $R_{\rm h}$, as a prime example. The gravitational
radius is an apparent horizon that separates null geodesics approaching the observer from
those that are receding. One could not argue that such a divided congruence of null geodesics
is consistent with homogeneity. But let us affirm that there is no conflict between these
two descriptions, because the universe is indeed homogeneous when viewed relative to the
comoving frame.

Given the thesis developed in this paper, it should be clear why the role of $R_{\rm h}$
is central to the manner in which numen quantum fluctuations were born and how they classicalized.
It is this gravitational horizon that delimited the size of the fluctuations, which
were isotropic, but nevertheless initially restricted in size to the Planck scale. Remember
that $\lambda_{\rm Pl}$ actually equals $R_{\rm h}$ at the time, $t_{\rm b}$, when mode
$k$ emerged into the semi-classical universe.

Crucially, an apparent horizon in the cosmological context is not an event horizon
\cite{Melia:2018c}. It may turn into one in our asymptotic future, but has not been static
up until today.  It has been growing, and causally-connected regions of spacetime continue to
change as $R_{\rm h}$ expands. The `measurement-like' process that created the numen fluctuations
at the Planck scale therefore also introduced an inhomogeneity relative to the observer who
appeared later, at time $t>t_{\rm b}$, outside the region bounded by $R_{\rm h}(t_{\rm b})$.
In the numen context, the largest CMB fluctuation we see today corresponds to $R_{\rm h}$
at decoupling. Since $\lambda_{\rm max}$ (corresponding to $k_{\rm min}$) grew at the same
rate as $R_{\rm h}$, this happens to be the largest-size mode created right after the Big
Bang, at the Planck time $t_{\rm Pl}$. This region has been exposed to us by the expanding
universe, so that we can now see the impact of that first quantum fluctuation on the CMB.
The seeding of numen fluctuations at the Planck scale not only created pure, single $k$-mode
fluctuations, but also explains why these quantum fluctuations have always been finite in
size, and why the universe we see is therefore a patchwork of inhomogeneities. At least
in this context, the breakdown of spherical symmetry was not due to some unknown process
in quantum mechanics itself. It was the result of an inherent property of the FLRW
spacetime when we take into account the physical characteristics of the gravitational
(or, apparent) horizon.

\subsection{Classical Phase Space Correlations}
A distinguishing feature of quantum fluctuations in the inflaton field is that their
frequency changed with time in response to the spacetime curvature they encountered as
they grew into the expanding universe. This led to the very strong time dependence
in $\Omega_{\bf k}$ (Eqn.~\ref{eqn:Omega}) which, according to the Wigner
function (Eqn.~\ref{eqn:squeeze}), then produced a highly squeezed state. For the
various reasons outlined in \S~2.2, however, it is not clear that this restructuring
of the dispersions about $u_{\bf k}$ and $\pi_{\bf k}$ is consistent with emerging
classical correlations in phase space.

The situation with numen quantum fluctuations was completely different, principally because
their frequency $\alpha_{\rm k}$ (Eqn.~\ref{eqn:freqnumen}) was constant in time. As we alluded
to earlier, this property results from the zero active mass equation-of-state ($\rho_\phi+
3p_\phi=0$), which produced a constant expansion rate with $a(t)\propto t$. The numen modes
therefore mimicked a classical harmonic oscillator.

\begin{figure}
\centering
\includegraphics[width=1.05\linewidth]{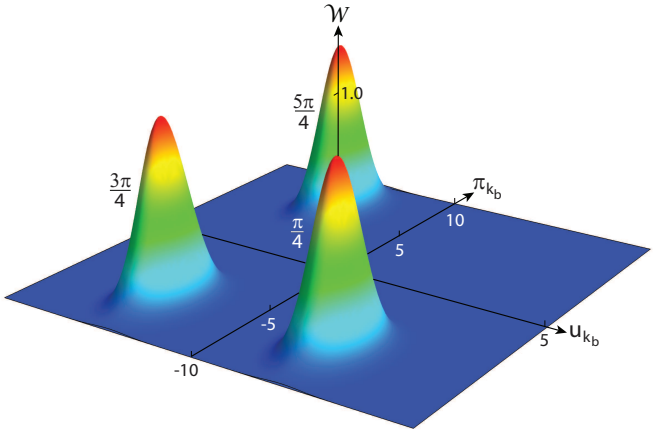}
\caption{Schematic diagram showing the Wigner function of the coherent numen state
(see Eqn.~\ref{eqn:final}). The state begins at the right, where $\alpha_{{\rm k}_{\rm b}}\tau=\pi/4$,
and moves clockwise about $(0,0)$. The second peak is plotted at $\alpha_{{\rm k}_{\rm b}}\tau=3\pi/4$,
and the third at $\alpha_{{\rm k}_{\rm b}}\tau=5\pi/4$. In generating these plots, the following
values were chosen for the constants in Equation~(\ref{eqn:final}): $\alpha_{{\rm k}_{\rm b}}=3$,
and $\mu=3$.}
\end{figure}

Such a coherent state is considered to be the most classical of all states because their Wigner
function peaks over tiny regions in phase space, and retains this configuration for all times.
Given a value of $u_{\bf k}$ for such a state, one may then obtain the corresponding canonical
momentum $\pi_{\bf k}$ very close to the value calculated using classical dynamics. The Wigner
function thus follows a classical trajectory, with minimal outward spread in all phase-space
directions.

To see this quantitatively, let us examine the Wigner function for a numen fluctuation,
\begin{equation}
{\mathcal{W}}_{\Psi_{{\rm k}_{\rm b}}}(u_{{\rm k}_{\rm b}},\pi_{{\rm k}_{\rm b}})\equiv
\int {dy\over \pi}\,\Psi_{{\rm k}_{\rm b}}^*(u_{{\rm k}_{\rm b}}+y)e^{2i\,\pi_{{\rm k}_{\rm b}}y}
\Psi_{{\rm k}_{\rm b}}(u_{{\rm k}_{\rm b}}-y)\;,
\end{equation}
which is reminiscent of the quantum harmonic oscillator. Its derivation is very well known,
so we won't dwell on the details (see, e.g., ref.~\cite{Case:2008}). With the 
ground-state solution to the Schr\"odinger Equation~(\ref{eqn:Schro}),
\begin{equation}
\Psi_{{\rm k}_{\rm b}}[u_{{\rm k}_{\rm b}}(\tau)]=\left({\alpha_{{\rm k}_{\rm b}}\over\pi}\right)^{1/4}
e^{-\alpha_{{\rm k}_{\rm b}}{u^2_{{\rm k}_{\rm b}}}/2}\;,
\label{eqn:psiground}
\end{equation}
we may easily evaluate the integral and find that
\begin{equation}
{\mathcal{W}}_{\Psi_{{\rm k}_{\rm b}}}(u_{{\rm k}_{\rm b}},\pi_{{\rm k}_{\rm b}})=
{1\over\pi}e^{-\alpha_{{\rm k}_{\rm b}}{u^2_{{\rm k}_{\rm b}}}-{\pi^2_{{\rm k}_{\rm b}}}/
\alpha_{{\rm k}_{\rm b}}}\;.
\end{equation}
To see how ${\mathcal{W}}_{\Psi_{{\rm k}_{\rm b}}}$ varies in time, let us briefly return
to Equations~(\ref{eqn:numenmode}) and (\ref{eqn:numennorm}), and solve for the time evolution
of $u_{{\rm k}_{\rm b}}(\tau)$ and $\pi_{{\rm k}_{\rm b}}(\tau)$ in terms of their values at
$\tau=0$: $u_{{\rm k}_{\rm b}}(0)$ and $\pi_{{\rm k}_{\rm b}}(0)$. We find that
\begin{eqnarray}
u_{{\rm k}_{\rm b}}(0)&=& u_{{\rm k}_{\rm b}}(\tau)\cos(\alpha_{{\rm k}_{\rm b}}\tau)-
{\pi_{{\rm k}_{\rm b}}\over\alpha_{{\rm k}_{\rm b}}}\sin(\alpha_{{\rm k}_{\rm b}}\tau) \nonumber\\
\pi_{{\rm k}_{\rm b}}(0)&=& \pi_{{\rm k}_{\rm b}}(\tau)\cos(\alpha_{{\rm k}_{\rm b}}\tau)+
\alpha_{{\rm k}_{\rm b}}u_{{\rm k}_{\rm b}}\sin(\alpha_{{\rm k}_{\rm b}}\tau)\;.\quad
\label{eqn:array}
\end{eqnarray}

Now take the Wigner function at time $\tau=0$ to represent the lowest energy state
for the field amplitude $u_{{\rm k}_{\rm b}}(0)$ shifted by a constant $\mu$. That is,
\begin{equation}
{\mathcal{W}}_{\Psi_{{\rm k}_{\rm b}}}(u_{{\rm k}_{\rm b}},\pi_{{\rm k}_{\rm b}},0)=
{1\over\pi}e^{-\alpha_{{\rm k}_{\rm b}}(u_{{\rm k}_{\rm b}}-\mu)^2-\pi^2_{{\rm k}_{\rm b}}/
\alpha_{{\rm k}_{\rm b}}}\;.
\end{equation}
\null\vskip 0in\noindent
With the use of Equation~(\ref{eqn:array}), this becomes
\begin{eqnarray}
{\mathcal{W}}_{\Psi_{{\rm k}_{\rm b}}}(u_{{\rm k}_{\rm b}},\pi_{{\rm k}_{\rm b}},\tau)&=&
{1\over\pi}e^{-\alpha_{{\rm k}_{\rm b}}[u_{{\rm k}_{\rm b}}(\tau)-
\mu\cos(\alpha_{{\rm k}_{\rm b}}\tau)]^2}\times\nonumber \\
&\null&\hskip-0.5in e^{-[\pi_{{\rm k}_{\rm b}}(\tau)+\alpha_{{\rm k}_{\rm b}}\mu
\sin(\alpha_{{\rm k}_{\rm b}}\tau)]^2/\alpha_{{\rm k}_{\rm b}}}\;.\quad
\label{eqn:final}
\end{eqnarray}
The Wigner function thus oscillates back and forth with an amplitude $\mu$ and angular
frequency $\alpha_{{\rm k}_{\rm b}}$---a coherent state with a very small dispersion
relative to the classicalized trajectory in the $(u_{{\rm k}_{\rm b}},\pi_{{\rm k}_{\rm b}})$
plane (see fig.~2).

This result has a rather simple interpretation. The fact that numen fluctuations were born
into the semi-classical universe in their ground state, with a time-independent frequency
$\alpha_{{\rm k}_{\rm b}}$, means that they functioned exactly like the coherent states of
a conventional quantum harmonic oscillator. They therefore acquired classical correlations
in the $(u_{{\rm k}_{\rm b}},\pi_{{\rm k}_{\rm b}})$ phase space from the very beginning.

\section{Conclusions}
The {\it Planck} data suggest that the primordial power spectrum has a cutoff, $k_{\rm min}$,
probably associated with the first quantum fluctuation to emerge out of the Planck domain
into the semi-classical universe. Moreover, this birth must have occurred at about the Planck
time---the earliest moment permitted by our current theories following the Big Bang. It is
easy to appreciate the physical significance of such an interpretation, because our current
(classical) theory of gravity is valid only down to the Planck scale, not beyond. It is
therefore tempting to view the manner in which these (numen) quantum fluctuations were born
as a physical consequence of the transition from quantum gravity to general relativity.

Insofar as the question of classicalization is concerned, this picture effectively removes
the hurdles faced by inflation. Leaving aside the related issue that the existence of
$k_{\rm min}$ challenges the viability of slow-roll inflation to simultaneously explain
the fluctuation spectrum and mitigate the horizon problem, which has been dealt with elsewhere
\cite{LiuMelia:2020}, one cannot ignore the challenge of explaining how fluctuations
in the inflaton field successfully transitioned from the quantum to classical domains.

With the interpretation we have examined in this paper, two factors stand out clearly
as the most influential. These are (i) the birth of quantum fluctuations with a fixed
length scale at the Planck time constitutes a `process' that effectively replaces the need
for a `measurement' in quantum mechanics; and (ii) the implied scalar field potential
has an equation-of-state consistent with the zero active mass condition in general
relativity. The spacetime associated with the expansion profile in this cosmology
would have allowed the fluctuations to emerge in their ground state, with a time-independent
frequency, $\alpha_{\rm k}$. This makes all the difference because the numen quantum
fluctuations were therefore essentially quantum harmonic oscillators, with classical
correlations of their canonical variables from the very beginning.

Should the proposal we have made in this paper turn out to be correct, an obvious question
concerns the need for inflation in cosmology. For a paradigm that has been with us for over
four decades, it is somewhat troubling that no complete theory yet exists. Some of the
complications appear to be that the data tend to shy away from its predictions as the
measurement precision continues to improve, rather than confirm the basic inflationary
premise with increasing confidence. The large-scale anomalies in the CMB anisotropies are
a good illustration of this point. Today, after three major satellite missions have
completed their work studying the CMB, one would reasonably have expected all of the
nuances associated with the primordial power spectrum ${\mathcal{P}}(k)$ to have
fallen into place. But they don't, and it is difficult to discount the observational
constraint that a zero $k_{\rm min}$ is now ruled out by the data at a relatively high
level of confidence \cite{MeliaLopez:2018,Melia:2020a}.

On the flip side, inflation may not even be necessary to solve the horizon problem,
which actually does not exist in cosmologies that avoided an early phase of deceleration
\cite{Melia:2013,Melia:2018d}. In that case, the fact that this paradigm struggles
explaining the angular correlation function in the CMB, and that its quantum fluctuations
have no obvious path to classicalization, may be indicators that it simply never happened.
The many observational tests already completed to examine this question would not reject
such a view (see, e.g., Table~2 in ref.~\cite{Melia:2018a}).

\section*{Acknowledgments}
I am grateful to the anonymous referee for an exceptional review of this 
manuscript and for suggesting several improvements to its presentation.


\begin{thebibliography}{00}
\bibitem{Starobinsky:1980} A. A. Starobinsky, PLB {\bf 91} (1980) 99
\bibitem{Guth:1981} A. H. Guth, PRD {\bf 23} (1981) 347
\bibitem{Linde:1982} A. D. Linde, PLB {\bf 108} (1982) 389
\bibitem{Albrecht:1982} A. Albrecht \& P. J. Steinhardt, PRL {\bf 48} (1982) 1220
\bibitem{Linde:1983} A. D. Linde, PLB {\bf 129} (1983) 177
\bibitem{Mukhanov:1981} V. F. Mukhanov \& G. Chibisov, JETP Lett. {\bf 33} (1981) 532
\bibitem{Mukhanov:1982} V. F. Mukhanov \& G. Chibisov, Sov. Phys. JETP {\bf 56} (1982) 258
\bibitem{Starobinsky:1982} A. A. Starobinsky, PLB {\bf 117} (1982) 175
\bibitem{Guth:1982} A. H. Guth \& S. Y. Pi, PRL {\bf 49} (1982) 1110
\bibitem{Hawking:1982} S. Hawking, PLB {\bf 115} (1982) 295
\bibitem{Bardeen:1983} J. M. Bardeen, P. J. Steinhardt \& M. S. Turner, PRD {\bf 28} (1983) 679
\bibitem{Stewart:1993} E. D. Stewart \& D. H. Lyth, PLB {\bf 302} (1993) 171
\bibitem{Martin:2006} J. Martin \& C. Ringeval, JCAP {\bf 0608} (2006) 009
\bibitem{Lorenz:2008} L. Lorenz, J. Martin \& C. Ringeval, JCAP {\bf 0804} (2008) 001
\bibitem{Larson:2011} D. Larson, J. Dunkley, G. Hinshaw, E. Komatsu, M. Nolta et al.,
Astrop. J. Suppl. {\bf 192} (2011) 16
\bibitem{Komatsu:2011} E. Komatsu et al., Astrop. J. Suppl. {\bf 192} (2011) 18
\bibitem{Melia:2020b} F. Melia, {\it The Cosmic Spacetime}. Taylor \& Francis, New York (2020)
\bibitem{Nambu:1992} Y. Nambu, PLB {\bf 276} (1992) 11
\bibitem{Mijic:1997} M. Mijic, IJMP-D {\bf 6} (1997) 505
\bibitem{Lee:2001} W.-L. Lee \& L.-Z. Fang, Europhys. Lett. {\bf 56} (2001) 904
\bibitem{Zurek:1981} W. H. Zurek, PRD {\bf 24} (1981) 1516
\bibitem{Joos:1985} E. Joos \& H. D. Zeh, Z. Phys. B {\bf 59} (1985) 223
\bibitem{Kiefer:1998} C. Kiefer, D. Polarski \& A. A. Starobinsky, Int. J. Mod. Phys. D {\bf 7} (1998) 455
\bibitem{Kiefer:2007} C. Kiefer, I. Lohmar, D. Plarski \& A. A. Starobinsky, CQG {\bf 24} (2007) 1699
\bibitem{Kiefer:2009} C. Kiefer \& D. Polarski, Adv. Sc. Lett. {\bf 2} (2009) 164
\bibitem{Adler:2003} S. L. Adler, Stud. Hist. Philos. Mod. Phys. {\bf 34} (2003) 135
\bibitem{Schlosshauer:2004} M. Schlosshauer, Rev. Mod. Phys. {\bf 76} (2004) 1267
\bibitem{Sudarsky:2011} D. Sudarsky, Int. J. Mod. Phys. D {\bf 20} (2011) 509
\bibitem{Bohm:1952a} D. Bohm, Phys. Rev. {\bf 85} (1952) 166
\bibitem{Bohm:1952b} D. Bohm, Phys. Rev. {\bf 85} (1952) 180
\bibitem{Planck:2018} Planck Collaboration, N. Aghanim, Y. Akrami et~al., arXiv e-prints, arXiv:1807.06209
(2018)
\bibitem{MeliaLopez:2018} F. Melia \& M. L{\'o}pez-Corredoira, Astron. Astrophys. {\bf 610} (2018) A87
\bibitem{Melia:2020a} F. Melia, Q. Ma, J.-J. Wei \& B. Yu, A\&A, submitted (2021)
\bibitem{Bardeen:1980} J. M. Bardeen, PRD {\bf 22} (1980) 1882
\bibitem{Kodama:1984} H. Kodama \& M. Sasaki, Prog. Theor. Phys. Suppl. {\bf 78} (1984) 1
\bibitem{Mukhanov:1992} V. F. Mukhanov, H. A. Feldman \& R. H. Brandenberger, Phys. Rep. {\bf 215}
(1992) 203
\bibitem{Bassett:2006} B. A. Bassett, S. Tsujikawa \& D. Wands, Rev. Mod. Phys. {\bf 78} (2006) 537
\bibitem{Bunch:1978} T. S. Bunch \& P.C.W. Davies, Proc. R. Soc. A {\bf 360} (1978) 117
\bibitem{Polarski:1996} D. Polarski \& A. A. Starobinsky, CQG {\bf 13} (1996) 377
\bibitem{Martin:2008} J. Martin, Lec. Notes in Phys. {\bf 738} (2008) 193
\bibitem{Halliwell:1985} J. J. Halliwell \& S. W. Hawking, PRD {\bf 31} (1985) 1777
\bibitem{Zurek:1992} W. H. Zurek, in Moscow 1990, Proceedings, Quantum gravity (192) 456
\bibitem{Brandenberger:1992} R. Brandenberger, H. Feldman \& V. Mukhavov, Phys. Rep. {\bf 215} (1992) 203
\bibitem{Laflamme:1993} R. Laflamme \& A. Matacz, Int. J. Mod. Phys. D {\bf 2} (1993) 171
\bibitem{Grishchuk:1997} L. P. Grishchuk \& J. Martin, PRD {\bf 56} (1997) 1924
\bibitem{Hartle:1998} J. B. Hartle, in Proceedings of the 11th Nishinomiya-Yukawa Symposium, ed. by
K. Kikkawa, H. Kunitomo \& H. Ohtsubo, World Scientific Singapore (1998)
\bibitem{Kiefer:2000} C. Kiefer, Nucl. Phys. Proc. Suppl. {\bf 88} (2000) 255
\bibitem{Castagnino:2003} M. Castagnino \& O. Lombardi, Int. J. Theor. Phys. {\bf 42} (2003) 1281
\bibitem{delCampo:2004} D. del Campo \& R. Parentani, PRD {\bf 70} (2004) 105020
\bibitem{Martin:2005} J. Martin, Lect. Notes Phys. {\bf 669} (2005) 199
\bibitem{Perez:2006} A. Perez, H. Sahlmann \& D. Sudarsky, Class. Q. Grav. {\bf 23} (2006) 2317
\bibitem{Grib:1999} A. A. Grib \& W. A. Rodrigues Jr., {\it Nonlocality in Quantum Physics}
(Kluwer Academic/Plenum Publishers) (1999)
\bibitem{Hartle:2006} J. B. Hartle, Int. J. Theor. Phys. {\bf 45} (2006) 1390
\bibitem{Hillery:1984} M. Hillery, R. F. O'Connell, M. O. Scully \& E. P. Wigner, Phys. Rep.
{\bf 106} (1984) 121
\bibitem{Lee:1995} H. Lee, Phys. Rep. {\bf 259} (1995) 150
\bibitem{Case:2008} W. B. Case, Am. J. Phys. {\bf 76} (2008) 937
\bibitem{Grishchuk:1990} L. Grishchuk \& Y. Sidorov, PRD {\bf 42} (1990) 3413
\bibitem{Colanero:2012} K. Colanero, eprint (arXiv:1208.0904) (2012)
\bibitem{Hinshaw:1996} G. Hinshaw, A.~J. Branday, C.~L. Bennett et~al., ApJ Letters {\bf 464} (1996) L25
\bibitem{Bennett:2003} C.~L. Bennett, R.~S. Hill, G. Hinshaw et~al., ApJ Supp. {\bf 148} (2003) 97
\bibitem{Bennett:2011} C.~L. Bennett, R.~S. Hill, G. Hinshaw et~al., ApJ Supp. {\bf 192} (2011) 17
\bibitem{Copi:2010} C.~J. Copi, D. Huterer, D.~J. Schwarz \& G.~D. Starkman, arXiv e-prints,
arXiv:1004.5602 (2010)
\bibitem{Melia:2019a} F. Melia, EPJ-C Letters {\bf 79} (2019) 455
\bibitem{LiuMelia:2020} J. Liu \& F. Melia, Proc. R. Soc. A {\bf 476} (2020) 20200364
\bibitem{Shafieloo:2004} A. Shafieloo \& T. Souradeep, PRD {\bf 70} (2004) 043523
\bibitem{Nicholson:2009} G. Nicholson \& C. R. Contaldi, JCAP {\bf 2009} (2009) 7
\bibitem{Hazra:2014} D. K. Hazra, A. Shafieloo \& T. Souradeep, JCAP {\bf 2014} (2014) 11
\bibitem{Ichiki:2010} K. Ichiki, R. Nagata \& J. Yokoyama, PRD {\bf 81} (2010) 083010
\bibitem{Tocchini:2005} D. Tocchini-Valentini, M. Douspis \& J. Silk, MNRAS {\bf 359} (2005) 31
\bibitem{Tocchini:2006} D. Tocchini-Valentini, Y. Hoffman \& J. Silk, MNRAS {\bf 367} (2006) 1095
\bibitem{Hunt:2014} P. Hunt \& S. Sarkar, JCAP {\bf 2014} (2014) 025
\bibitem{Hunt:2015} P. Hunt \& S. Sarkar, JCAP {\bf 2015} (2015) 052
\bibitem{Planck:2016} Planck Collaboration, A\&A {\bf 594} (2016) A20
\bibitem{Sachs:1967} R.~K. Sachs \& A.~M. Wolfe, ApJ {\bf 147} (1967) 73
\bibitem{Destri:2008} C. Destri, H.~J. de Vega \& N.~G. Sanchez, PRD {\bf 78} (2008) 023013
\bibitem{Destri:2010} C. Destri, H.~J. de Vega \& N.~G. Sanchez, PRD {\bf 81} (2010) 063520
\bibitem{Melia:2018a} F. Melia, MNRAS {\bf 481} (2018) 4855
\bibitem{Melia:2013} F. Melia, Astron. Astrophys. {\bf 553} (2013) id. A76
\bibitem{Melia:2018b} F. Melia, EPJ-C Lett. {\bf 78} (2018) 739
\bibitem{Melia:2019b} F. Melia, Annals Phys. {\bf 411} (2019) 167997
\bibitem{Abbott:1984} L. F. Abbott \& M. B. Wise, Nucl. Phys. {\bf 244} (1984) 541
\bibitem{Lucchin:1985} F. Lucchin \& S. Materrese, PRD {\bf 32} (1985) 1316
\bibitem{Barrow:1987} J. Barrow, PLB {\bf 187} (1987) 12
\bibitem{Liddle:1989} A. R. Liddle, PLB {\bf 220} (1989) 502
\bibitem{Melia:2018c} F. Melia, Am. J. Phys. {\bf 86} (2018) 585
\bibitem{Martin:2001} J. Martin \& R. H. Brandenberger, PRD {\bf 63} (2001) 123501
\bibitem{Hollands:2002} S. Hollands \& R. M. Wald, Gen. Relativ. Gravit. {\bf 34} (2002) 2043
\bibitem{Brandenberger:2002} R. H. Brandenberger \& P. M. Ho, PRD {\bf 66} (2002) 023517
\bibitem{Hassan:2003} S. F. Hassan \& M. S. Sloth, Nucl. Phys. B {\bf 674} (2003) 434
\bibitem{Brouzakis:2012} N. Brouzakis, J. Rizos \& N. Tetradis, PLB {\bf 708} (2012) 170
\bibitem{Dvali:2012} G. Dvali \& C. Gomez, JCAP {\bf 1207} (2012) 015
\bibitem{Weinberg:1972} W. Weinberg, {\it Gravitation And Cosmology: Principles And Applications
Of The General Theory Of Relativity}. Wiley, New York (1972)
\bibitem{Melia:2018d} F. Melia, EPJ-C Lett. {\bf 78} (2018) 739
\end{thebibliography}
\end{document}